\documentclass{svproc}
\usepackage{blindtext}
\usepackage{booktabs}
\usepackage{multirow}
\usepackage{siunitx}
\usepackage{graphicx}
\usepackage[utf8]{inputenc}
\usepackage{graphics} 
\usepackage{epsfig} 
\usepackage{amsmath} 
\usepackage{amssymb}  
\usepackage{multicol}
\usepackage{mathtools, cuted}
\usepackage[cmintegrals]{newtxmath}
\usepackage{epstopdf}
\usepackage{graphics} 
\usepackage{epsfig} 

\usepackage{cite}
\usepackage{multicol}
\usepackage{mathtools, cuted}
\usepackage[cmintegrals]{newtxmath}
\usepackage{mathtools}

\usepackage{helvet}
\usepackage{courier}
\usepackage{subfigure}
\usepackage{type1cm}
\usepackage{epsfig} 
\usepackage{amsmath} 
\usepackage{amssymb}  
\usepackage{cite}
\usepackage{multicol}
\usepackage{mathtools, cuted}
\usepackage[cmintegrals]{newtxmath}
\usepackage{makeidx}         
\usepackage{algorithmic}

\usepackage{multicol}        
\usepackage[bottom]{footmisc}
\usepackage{caption}
\usepackage{nomencl}
\usepackage{algorithmic}
\usepackage{multicol}        
\usepackage[bottom]{footmisc}
\usepackage{caption}
\usepackage{nomencl}

\newtheorem{assumption}{Assumption}
\usepackage{url}
\usepackage{graphicx}
\usepackage{multicol}
\usepackage{footmisc}
\usepackage[numbers]{natbib}

\begin{document}
\mainmatter              
\title{Drones Practicing Mechanics}
\titlerunning{Teacher Drones}  
%
\author{Harshvardhan Uppaluru \and Hossein Rastgoftar}
\authorrunning{Uppaluru and Rastgoftar} 
%
\tocauthor{Harshvardhan Uppaluru, Hossein Rastgoftar}
\institute{Aerospace and Mechanical Engineering Department, \\ University of Arizona, Tucson, AZ, USA,\\
\email{\{huppaluru, hrastgoftar\} @email.arizona.edu}\\
}

\maketitle              

\begin{abstract}
Mechanics of materials is a classic course of engineering presenting the fundamentals of strain and stress analysis to junior undergraduate students in several engineering majors. So far, material deformation and strain have been only analyzed using theoretical and numerical approaches, and they have been experimentally validated by expensive machines and tools. This paper presents a novel approach for strain and deformation analysis by using quadcopters. We propose to treat quadcopters as finite number of particles of a deformable body and apply the principles of continuum mechanics to illustrate the concept of axial and shear deformation by using quadcopter hardware in a $3$-D motion space. The outcome of this work can have  significant impact on undergraduate education by  filling the gap between in-class learning and hardware realization and experiments, where we introduce new roles for drones as ``teachers'' providing a great opportunity for practicing theoretical concepts of mechanics in a fruitful and understandable way.



\end{abstract}

\section{Introduction}

An active area of research in multi-agent systems includes coordination, formation and cooperative control with a wide range of applications such as surveillance \cite{allouche2010multi}, search and rescue missions \cite{kleiner2013rmasbench}, precision agriculture \cite{ali2010multi}, structural health monitoring \cite{yuan2005distributed}, air traffic monitoring \cite{idris2018air}. Virtual structure \cite{ren2002virtual, low2011flexible}, consensus control \cite{ren2007information, cao2015leader, shao2018leader}, containment control \cite{li2016containment, zhao2015robust, notarstefano2011containment} and continuum deformation \cite{rastgoftar2016continuum} are some of the prevalent and existing methodologies for multi-agent system coordination that have been widely investigated. 

\subsection{Related Work}

The virtual structure \cite{lewis1997high, young2001control, beard2000feedback} approach is generally utilized as a centralized coordination method where the multi-agent formation is represented as a single structure and a rigid body. Given some orientation, the virtual structure advances in a certain direction as a rigid body while preserving the rigid geometric relationship between multiple vehicles. 
Consensus control is a decentralized multi-agent coordination approach with several multi-agent coordination applications proposed such as leaderless multi-agent consensus \cite{qin2016leaderless, ding2019leaderless} and leader-follower consensus \cite{wu2018leader}. Fixed communication topology and switching inter-agent communication are other areas under multi-agent consensus previously investigated \cite{wang2018fixed, wen2016group}. Stability of the consensus control in the presence of communication delays was also studied \cite{zhou2018h, zhang2019delay}. 

Containment control is a decentralized leader-follower method where a finite number of leaders guide the followers through local communication. Containment control of a multi-agent system in finite-time was considered \cite{wang2013distributed, liu2015distributed}. Necessary and sufficient conditions for containment control stability and convergence were established in \cite{cao2012distributed, ji2008containment}. The authors in \cite{cao2012distributed, notarstefano2011containment, li2015containment} explored containment control under fixed and switching inter-agent communication. Containment control under the presence of time-varying delays affecting multi-agent coordination was analyzed in \cite{shen2016containment, liu2014containment}. 

Continuum deformation is another decentralized multiagent coordination approach that treats agents as particles of a continuum, deforming in a $3$-D space. An $n$-D continuum deformation coordination has $n+1$ leaders, located at the vertices of an $n$-D simplex at any time $t$. Leaders plan desired trajectories and followers obtained the trajectories through local communication \cite{rastgoftar2016continuum}. Though, continuum deformation and containment control are similar and both are decentralized leader-follower methods, continuum deformation ensures inter-agent collision avoidance, obstacle collision avoidance and agent containment by formally specifying and verifying safety in a large-scale agent coordination system \cite{ rastgoftar2018safe, rastgoftar2019safe}. In an obstacle-laden environment, a large scale multi-agent system can safely and aggressively deform using continuum deformation coordination. \cite{romano2019experimental} experimentally evaluated continuum deformation coordination in $2$D with a team of $5$ quadcopters. 

\subsection{Contributions and Outline}
In this paper, we show how small quadcopters treated as particles of a deformable body can practice linear deformation in a $3$-D motion space. By assigning lower bound on axial strains of the desired continuum deformation coordination, we experimentally validate inter-agent collision avoidance in an aggressive continuum deformation coordination. While numerical and analytic approaches are available for analyzing material deformation, our work proposes a new approach for analyzing material deformation and strain by using quadcopter hardware. This will provide a great opportunity for integration of research into education by devising new approaches for teaching the core concepts of mechanics in a fruitful and understandable manner.

This paper is organized as follows: Basics of linear deformation is presented in Section \ref{Prelim}. Our approach for hardware realization of the continuum deformation is detailed in Section \ref{Methodology}. The experimental validation results are presented in Section \ref{Experimental Setup} and followed by Conclusion in Section \ref{Conclusion}.

\section{Preliminaries}\label{Prelim}
We consider liner transformation of deformable bodies specified by a homogeneous transformation in a $3$-D motion space, given by
\begin{equation}
\label{mainraw}
    \mathbf{r}_i(t)=\mathbf{Q}(t)\mathbf{r}_{i,0}+\mathbf{d}(t),\qquad t\geq t_0,
\end{equation}
In Eq. \eqref{mainraw}, $t_0$ is the initial time, $t\geq t_0$ is the current time,  $\mathbf{r}_{i,0}\in \mathbb{R}^3$ is the material position of particle $i$, $\mathbf{d}(t)$ is the rigid-body displacement vector, $\mathbf{r}_i(t)$ is the current desired position of particle $i$, and $\mathbf{Q}(t)$ is the Jacobian matrix that can be decomposed as follows:
\begin{equation}\label{matrixQ}
    \mathbf{Q}(t)=\mathbf{R}\left(t\right)\mathbf{E}\left(t\right)
\end{equation}
In Eq. \eqref{matrixQ}, $\mathbf{R}(t)$ is an orthogonal rotation matrix and 
\begin{equation}
\label{puredeformation}
\begin{split}
    \mathbf{E}\left(t\right)=&\begin{bmatrix}
    \epsilon_{xx}\left(t\right)&\epsilon_{xy}\left(t\right)&\epsilon_{xz}\left(t\right)\\
    \epsilon_{xy}\left(t\right)&\epsilon_{yy}\left(t\right)&\epsilon_{yz}\left(t\right)\\
    \epsilon_{xz}\left(t\right)&\epsilon_{yz}\left(t\right)&\epsilon_{zz}\left(t\right)\\
    \end{bmatrix}
\end{split}
\end{equation}
is a positive definite strain matrix. 
\begin{assumption}\label{assum1}
We assume that the material configuration of the continuum is the same as the initial configuration. Therefore, material position $\mathbf{r}_{i,0}=\mathbf{r}_i(t_0)$ is the same as the initial position $\mathbf{r}_i(t_0)$ for every material particle $i$.
\end{assumption}

\begin{assumption}\label{assum2}
We assume that the $\mathbf{d}\left(t_0\right)=0$.
\end{assumption}

By considering Assumptions \ref{assum1} and \ref{assum2}, $\mathbf{Q}$ becomes the identity matrix at the initial time $t_0$, i.e., $\mathbf{Q}\left(t_0\right)=\mathbf{I}_3$.

\section{Methodology}\label{Methodology}

We consider $N$ quadcopters coordinating in a $3$-D motion space where they are identified by set $\mathcal{V}=\left\{1,\cdots,N\right\}$.
We define quadcopters as particles of a deformable body and    let Eq. \eqref{mainraw} define the desired continuum deformation of the quadcopter team. Without loss of generality, we only focus on realization of pure deformation, thus, we set  $\mathbf{d}(t)=\mathbf{0}\in\mathbb{R}^{3\times 1}$ and $\mathbf{R}=\mathbf{I}\in \mathbb{R}^{3\times 3}$ at any time $t$ (i.e rigid-body displace and rotation are both zero at any time $t$). Under this assumption, the quadcopter team continuum deformation, given by \eqref{mainraw}, simplifies to
\begin{equation}
    \label{mainsimple}
    \mathbf{r}_i(t)=\mathbf{E}\left(t\right)\mathbf{r}_{i,0},\qquad \forall i\in \mathcal{V},~ \forall t\in \left[t_0,t_f\right]
\end{equation}
where $t_0$ and $t_f$ denote the initial and final times, $\mathbf{r}_{i,0}$ and $\mathbf{r}_i(t)$ are the material and the current desired position of quacopter $i\in \mathcal{V}$, and positive definite strain matrix  $\mathbf{E}\left(t\right)$, defined by \eqref{puredeformation}, specifies the axial strains and shear deformations in a liner deformation scenario. \textit{While particles have infinitesimal size in a material deformation, here, quadcopters, treated as particles of a deformable body, are rigid and cannot deform. Therefore, realization of  linear deformation by  a quadcopter team requires to assure inter-agent collision avoidance via constraining the lower bound of the principal strains at any time $t$.} To this end, the principal strains, defined as eigenvalues of matrix $\mathbf{E}$, must all be greater than $\epsilon_{min}$ where $\epsilon_{min}$ is obtained based on (i) quadcopter size, (ii) quadcopter trajectory control performance, and (iii) the minimum separation distance between every two quadcopters in the initial (material) configuration \cite{rastgoftar2016continuum}. To assign $\epsilon_{min}$, we will make the following assumptions:
\begin{assumption}\label{assum3}
Every quadcopter can be enclosed by a ball of radius $r$.
\end{assumption}
\begin{assumption}\label{assum4}
The trajectory tracking error of every quadcopter is less than $\delta$. Therefore, 
\begin{equation}
    \bigwedge_{i\in \mathcal{V}}\left(\|\mathbf{p}_i(t)-\mathbf{r}_i(t)\|\leq \delta \right),\qquad \forall t\in \left[t_0,t_f\right]
\end{equation}
where $\mathbf{p}_i(t)$ is the actual position of quadcopter $i\in \mathcal{V}$ at time $t\in \left[t_0,t_f\right]$. 
\end{assumption}
By imposing Assumption \ref{assum3} and \ref{assum4}, we obtain
\begin{equation} \label{eq:minimum separation distance}
    \epsilon_{min}=\dfrac{2\left(\delta+r\right)}{p_{min}}
\end{equation}
where $p_{min}$ is the minimum separation distance between every two quadcopters in the initial (reference) configuration \cite{rastgoftar2016continuum}. 

\begin{figure}[ht]
\includegraphics[width=\textwidth]{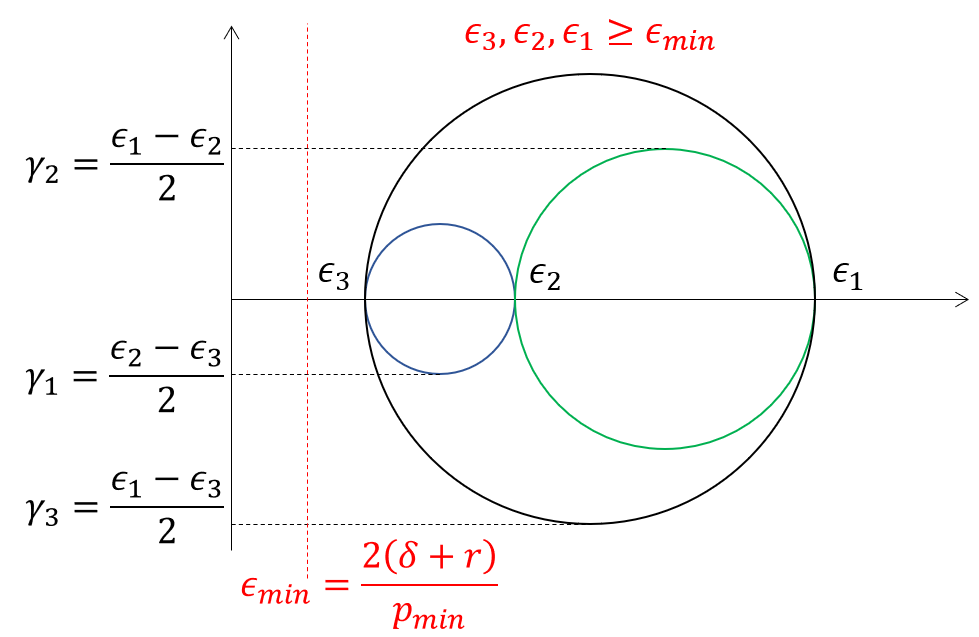}
\centering
\caption{Graphical representation of axial and shear strains by using Mohr circle. }
\label{MohrCircle}
\end{figure}

To formally characterize safety, we need to decompose matrix $\mathbf{E}$ and define it based on axial and shear strains. To this end,
we use the $3-2-1$ Euler angle standard to specify rigid-body rotation in a $3$-D motion space by matrix
\[
\resizebox{0.99\hsize}{!}{%
$
\mathbf{L}_{\mathrm{Euler}}\left(x_1,x_2,x_3\right)= \begin{bmatrix}
    \cos{x_2} \cos{x_3}&\cos{x_2} \sin{x_3} &-\sin{x_2}\\
  \sin{x_1}\sin{x_2} \cos{x_3}-\cos{x_1}\sin{x_3}&\sin{x_1}\sin{x_2} \sin{x_3}+\cos{x_1}\cos{x_3}&\sin{x_1}\cos{x_2}\\
  \cos{x_1}\sin{x_2} \cos{x_3}+\sin{x_1}\sin{x_3} &\cos{x_1}\sin{x_2} \sin{x_3}-\sin{x_1}\cos{x_3}&\cos{x_1}\cos{x_2}
\end{bmatrix}
,
$
}
\]
where $x_1$, $x_2$, and $x_3$ are the first, second, and third Euler angles, respectively. Matrix $\mathbf{E}(t)$ is decomposed as 
\begin{equation}\label{mainEEEE}
    \mathbf{E}\left(t\right)=\mathbf{L}_{\mathrm{Euler}}^T\left(\phi_d(t),\theta_d(t),\psi_d(t)\right)\mathbf{diag}\left(\epsilon_1(t),\epsilon_2(t),\epsilon_3(t)\right)\mathbf{L}_{\mathrm{Euler}}\left(\phi_d(t),\theta_d(t),\psi_d(t)\right)
\end{equation}
at any time $t\in \left[t_0,t_f\right]$, where $\phi_d(t)$, $\theta_d(t)$, $\psi_d(t)$ are the shear deformation angles and $\epsilon_1(t)$, $\epsilon_2(t)$, $\epsilon_3(t)$. The axial and shear strains specified by matrix $\mathbf{E}\left(t\right)$ can be graphically illustrated by using the Mohr circle, as shown in Fig. \ref{MohrCircle}. By using the Mohr circle, we can assign the bounds on the deformation angles $\phi_d(t)$, $\theta_d(t)$, and $\psi_d(t)$ that will improve safety of the continuum deformation coordination.

\begin{figure}[ht]
\includegraphics[width=\textwidth]{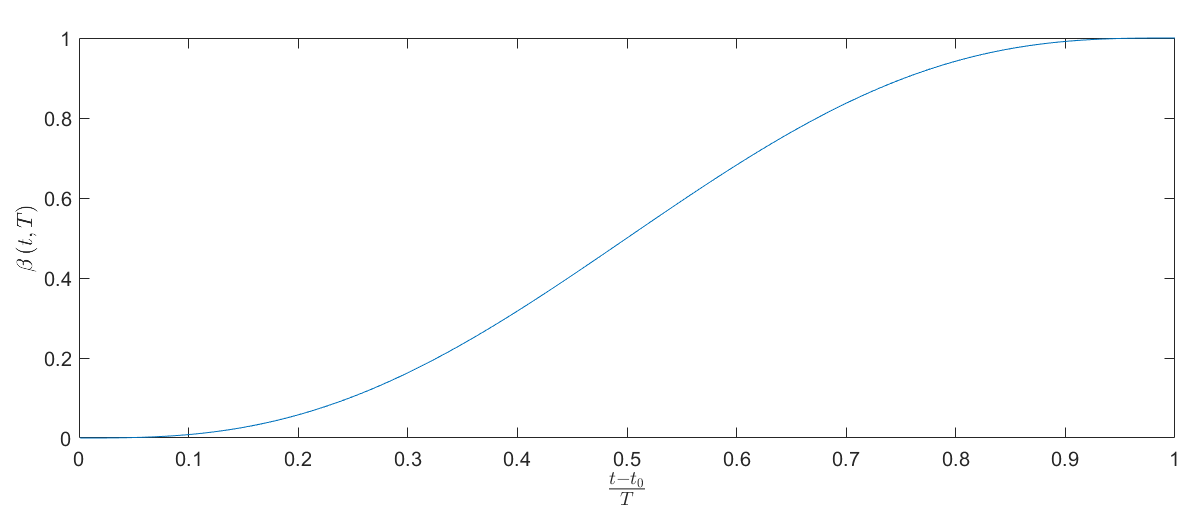}
\centering
\caption{$\beta\left(t,T\right)$ versus ${\frac{t-t_0}{T}}$ for ${\frac{t-t_0}{T}}\in \left[0,1\right]$. }
\label{beta}
\end{figure}

\begin{figure}[ht]
\includegraphics[width=\textwidth]{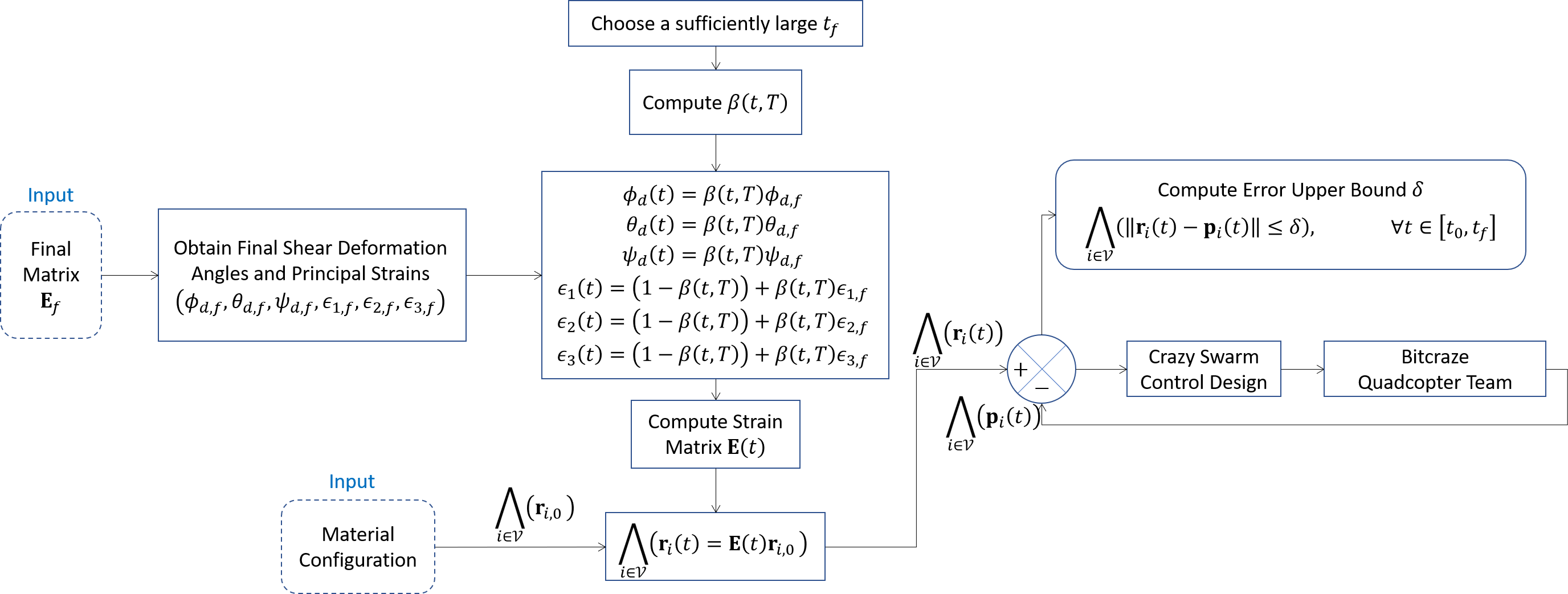}
\centering
\caption{The block diagram of quadcopter team coordination planning and control. }
\label{blockdiagram}
\end{figure}

For quadcopter team continuum deformation coordination, positive definite matrix $\mathbf{E}_f=\mathbf{E}\left(t_f\right)=\left[\epsilon_{ij}\left(t_f\right)\right]$ is known but final time $t_f$ is free. The problem of continuum deformation planning consists of (i) assignment of final time $t_f$ and (ii) obtaining matrix $\mathbf{E}(t)$ for $t\in \left[t_0,t_f\right]$ by performing the following two steps:
\begin{itemize}
    \item \textbf{Step 1--Choosing Final Time $t_f$:} The final time $t_f$
 needs to be sufficiently large such that the principal strains, denoted by $\epsilon_1$, $\epsilon_2$, and $\epsilon_3$, satisfy the following safety constraint:
\begin{equation}\label{MainSafetyy}
    \bigwedge_{i=1}^3\left(\epsilon_{min}\leq \epsilon_i(t)\right),\qquad \forall t\in\left[t_0,t_f\right].
\end{equation}
In Ref. \cite{rastgoftar2021safe}, it was shown how the minimum final time $t_f^*$ can be obtained by using bi-section method such that all satisfied. 
    \item \textbf{Step 2--Specifying $\mathbf{E}\left(t\right)$ for $t\in \left[t_0,t_f\right]$:}
To assure that the safety constraint \eqref{MainSafetyy} is satisfied at any time $t\in \left[t_0,t_f\right]$, we need to decompose matrix $\mathbf{E}\left(t\right)$ and  perform the following sub-steps to plan $\mathbf{E}(t)$ (for $t\in \left[t_0,t_f\right]$):
\begin{itemize}
    \item \textbf{A. Assignment of Final Shear Deformation Angles and Ultimate Principal Strains:} Given $\mathbf{E}_f$, we first obtain the final values of the shear deformation angles, denoted by $\phi_{d,f}=\phi_d(t_f)$, $\theta_{d,f}=\theta_d(t_f)$, and $\psi_{d,f}=\psi_d(t_f)$, and the ultimate principal strain values, denoted by $\epsilon_{1,f}=\epsilon_1(t_f)$, $\epsilon_{2,f}=\epsilon_2(t_f)$, and $\epsilon_{3,f}=\epsilon_3(t_f)$, by solving six nonlinear equations provided by Eq. \eqref{mainEEEE}.
    \item \textbf{B. Planning of the Shear Deformation Angles and Principal Strains at Every Time $t\in \left[t_0,t_f\right]$:} We first define the fifth order polynomial 
    \begin{equation}
        \beta\left(t,T\right)=15\left({\frac{t-t_0}{T}}\right)^5-16\left({\frac{t-t_0}{T}}\right)^4+10\left({\frac{t-t_0}{T}}\right)^3,\qquad t\in \left[t_0,t_f\right]
    , 
    \end{equation}
    where $T=t_f-t_0$ is the travel time, ${\beta}\left(0,T\right)=0$, $\dot{\beta}\left(0,T\right)=\dot{\beta}\left(t_f,T\right)=0$, $\ddot{\beta}\left(0,T\right)=\ddot{\beta}\left(t_f,T\right)=0$, and ${\beta}\left(t_f,T\right)=1$ (we plotted ${\beta}\left(t,T\right)$ versus ${\frac{t-t_0}{T}}$ in Fig. \ref{beta}). Then, the shear deformation angles and pincipal strains are defined by
    \begin{subequations}
    \begin{equation}
        \phi_d\left(t\right)=\phi_{d,0}\left(1-\beta\left(t,T\right)\right)+\phi_{d,f}\beta\left(t,T\right),\qquad \forall t\in \left[t_0,t_f\right],
    \end{equation}
        \begin{equation}
        \theta_d\left(t\right)=\theta_{d,0}\left(1-\beta\left(t,T\right)\right)+\theta_{d,f}\beta\left(t,T\right),\qquad \forall t\in \left[t_0,t_f\right],
    \end{equation}
            \begin{equation}
        \psi_d\left(t\right)=\psi_{d,0}\left(1-\beta\left(t,T\right)\right)+\psi_{d,f}\beta\left(t,T\right),\qquad \forall t\in \left[t_0,t_f\right],
    \end{equation}
            \begin{equation}
        \epsilon_1\left(t\right)=\epsilon_{1,0}\left(1-\beta\left(t,T\right)\right)+\epsilon_{1,f}\beta\left(t,T\right),\qquad \forall t\in \left[t_0,t_f\right],
    \end{equation}
                \begin{equation}
        \epsilon_2\left(t\right)=\epsilon_{2,0}\left(1-\beta\left(t,T\right)\right)+\epsilon_{2,f}\beta\left(t,T\right),\qquad \forall t\in \left[t_0,t_f\right],
    \end{equation}
                \begin{equation}
        \epsilon_3\left(t\right)=\epsilon_{3,0}\left(1-\beta\left(t,T\right)\right)+\epsilon_{3,f}\beta\left(t,T\right),\qquad \forall t\in \left[t_0,t_f\right],
    \end{equation}
    \end{subequations}
   where  $\phi_{d,0}=\phi_d\left(t_0\right)=0$, $\theta_{d,0}=\theta_d\left(t_0\right)=0$, $\psi_{d,0}=\psi_d\left(t_0\right)=0$, $\epsilon_{1,0}=\epsilon_1\left(t_0\right)=1$, $\epsilon_{2,0}=\epsilon_2\left(t_0\right)=1$, and $\epsilon_{3,0}=\epsilon_3\left(t_0\right)=1$ since $\mathbf{E}\left(t_0\right)=\mathbf{I}\in \mathbb{R}^{3\times 3}$ is an identity matrix (see Assumptions \ref{assum1} and \ref{assum2}).
\item \textbf{C. Assignment of Matrix $\mathbf{E}\left(t\right)$:} By knowing the shear deformation angles and principal strains at any time $t\in \left[t_0,t_f\right]$, matrix $\mathbf{E}\left(t\right)$ is assigned by using Eq. \eqref{mainEEEE} at any time $t\in \left[t_0,t_f\right]$.
\end{itemize}

\end{itemize}
Functionality of our proposed approach for experimental evaluation of continuum deformation coordination is shown in Fig. \ref{blockdiagram}.

\section{Experiments}

In this section, we will first explain the hardware, software, and environment configurations that have been utilized for these experiments (Subsection \ref{Experimental Setup}). In subsection \ref{Experimental Evaluation}, we provide the findings of an experimental evaluation of the quadcopter team's 2-D and 3-D continuum deformation coordination.

\subsection{Experimental Setup}
\label{Experimental Setup}

\begin{figure}[ht]
\centering
\subfigure[]{\includegraphics[height=4cm, width=.24\linewidth]{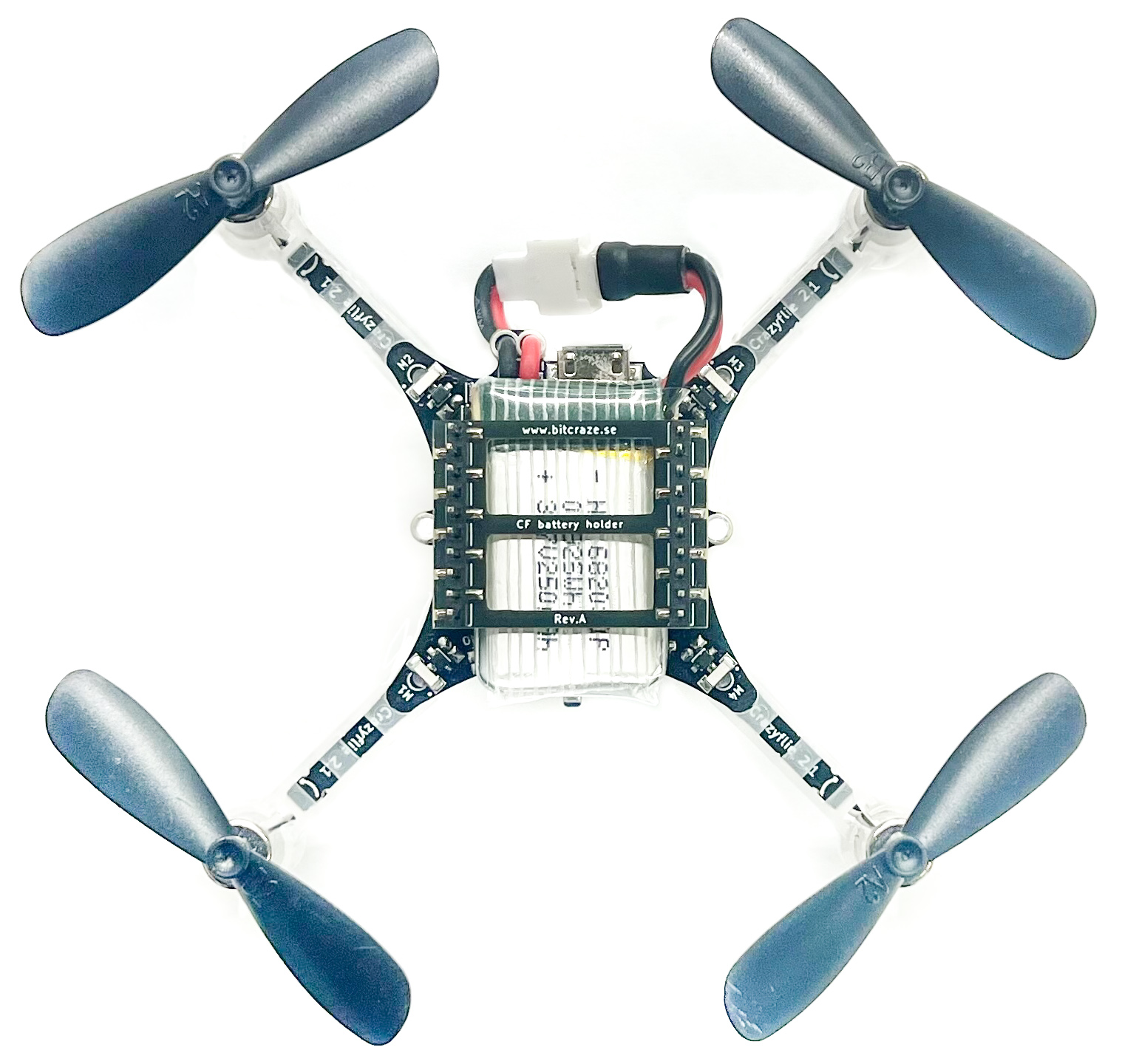}}
\subfigure[]{\includegraphics[height=4cm, width=.24\linewidth]{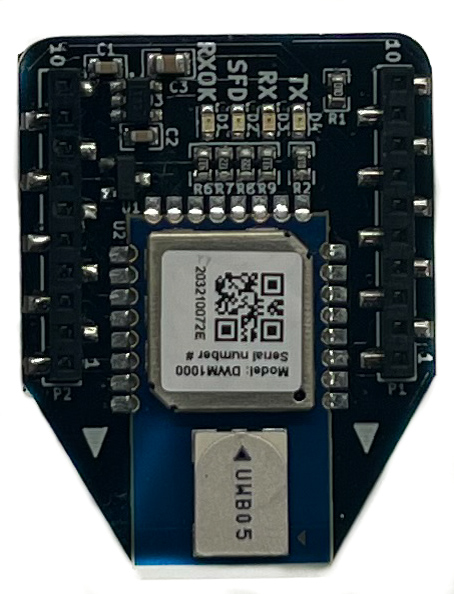}}
\subfigure[]{\includegraphics[height=4cm, width=.24\linewidth]{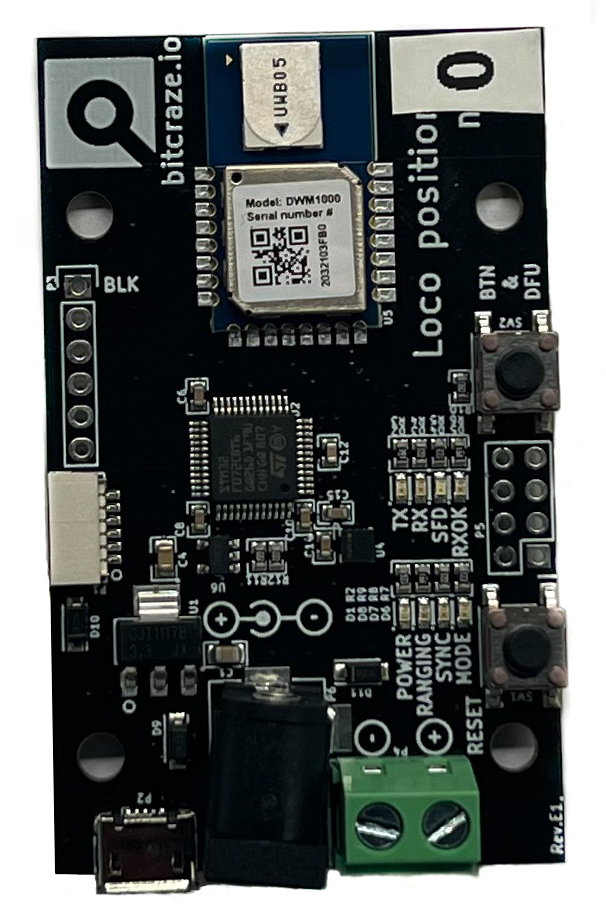}}
\subfigure[]{\includegraphics[height=4cm, width=.24\linewidth]{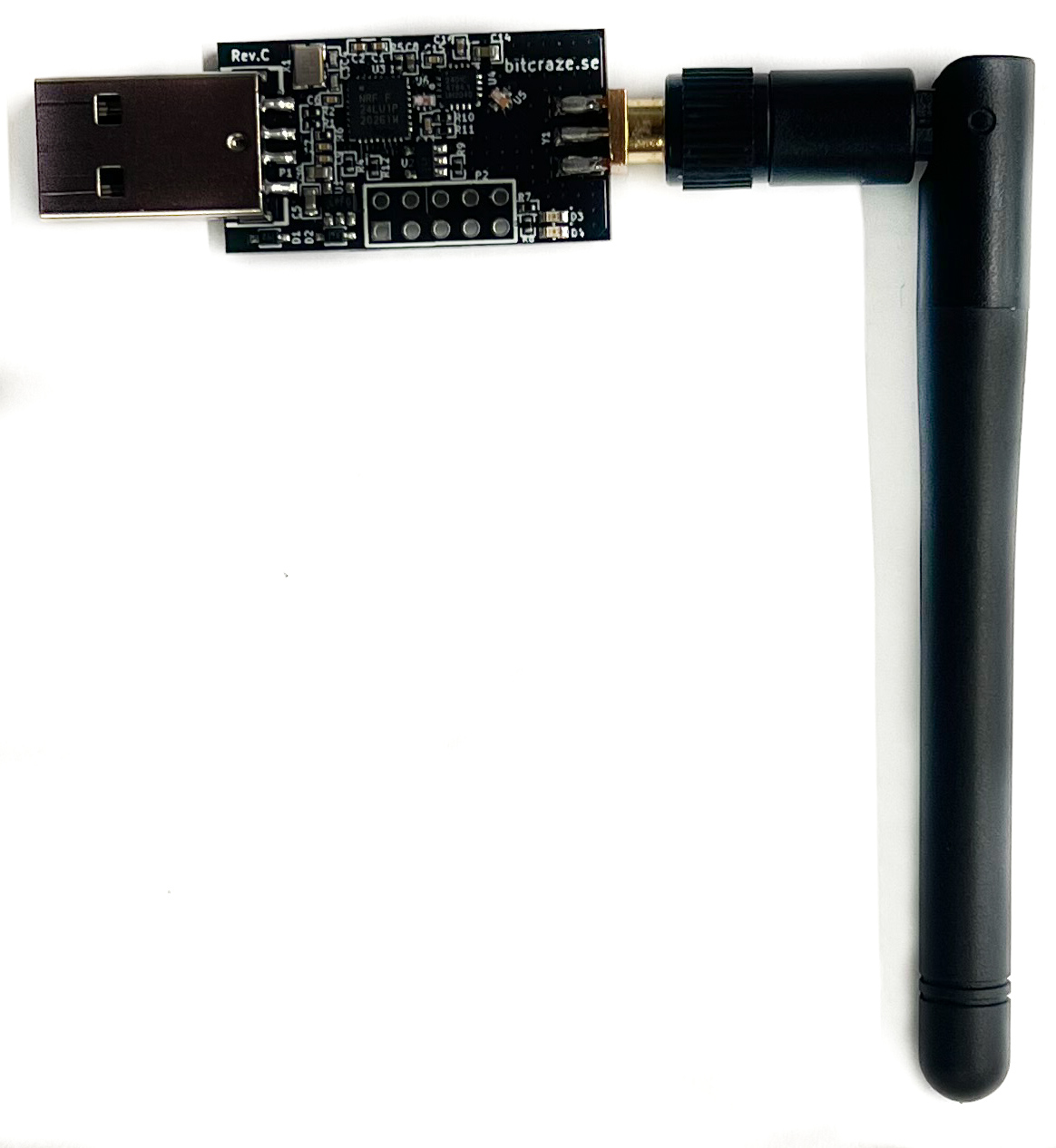}}
\caption{(a) Crazyflie quadcopters used in our experiments (b) Loco Positioning Deck (c) Loco Positioning Node (d) Crazyradio PA}
\label{fig:devices}
\end{figure}

The proposed quadcopter team continuum deformation coordination was evaluated on a hardware configuration comprised of Crazyflie 2.1\footnote{\url{https://www.bitcraze.io/products/crazyflie-2-1/}}, Bitcraze's\footnote{\url{https://www.bitcraze.io/}} open-source, open-hardware nano quadcopter. Four coreless DC motors and 45 mm plastic propellers are included with each Crazyflie. As a result, the quadcopter is just 92mm diagonal rotor-to-rotor, 29mm tall, and weighs around 27g with the battery, making it ideal for dense formation flight.

The  Crazyflie quadcopter is equipped with two microcontroller units (MCUs): a Cortex-M4, 168MHz, 192kb SRAM, 1Mb flash primary controller, STM32F405, and a Cortex-M0, 32MHz, 16kb SRAM, 128kb flash controller, nRF5182. The primary application MCU is STM32F405, which is capable of inertial state estimation and control tasks, while nRF51822 is responsible for radio and power management. Each Crazyflie can fly for up to 7 minutes on a single charge of its 240mAh LiPo battery.

Each Crazyflie in our system has a Loco Positioning deck to expand its onboard possibilities. A Loco Positioning Deck serves as a tag in Bitcraze's Ultra Wide Band radio-based Loco Positioning System (LPS)\footnote{\url{https://www.bitcraze.io/documentation/system/positioning/loco-positioning-system/}}. In the flying space, a number of Anchors, also known as Loco Positioning Nodes, serve as reference points. A two-way transmission of brief high-frequency radio communications between Anchors and Tags enables the system to determine Tag position in 3D space. This technology, which functions similarly to an indoor GPS system, is used to determine the absolute 3D position of a quadcopter in space with decimeter-level precision. LPS provides a variety of positioning modes, including Two Way Ranging (TWR), Time Difference of Arrival 2 (TDoA2), and Time Difference of Arrival 3 (TDoA3). We choose the Time Difference of Arrival 3 (TDoA3) positioning mode since it can accommodate an unlimited number of Crazyflies and Anchors. Each Crazyflie connects with a PC through the Crazyradio PA\footnote{\url{https://www.bitcraze.io/products/crazyradio-pa/}}, a long range 2.4 GHz USB radio capable of sending up to two megabits per second in 32-byte packets.

All simulations were performed using MATLAB on a laptop running Ubuntu 20.04 with an Intel Core i7-10610U 1.8 GHz CPU, Mesa Intel UHD Graphics card, and 16 GB of RAM. We employ ROS Noetic in conjunction with the Crazyswarm ROS stack built by the USC-ACT lab\footnote{\url{https://crazyswarm.readthedocs.io/en/latest/}} for flight experiments. All preliminary flight tests were carried out at the University of Arizona's Scalable Move and Resilient Transversability (SMART) lab's indoor flying area. The environmental setup for tests consists of a total of eight Loco Positioning Nodes positioned at the corners of the flying volume of size $5.5 \times 4.5 \times 2 m^3$.

\subsection{Experimental Evaluation}\label{Experimental Evaluation}

To ensure inter-agent collision avoidance, the minimum separation distance in the initial configuration should be large enough such that the safety condition  \eqref{eq:minimum separation distance} is satisfied. According to Bitcraze, the tracking of quadcopters is accurate up to $0.1m$ when the Loco Positioning System (LPS) is employed. To be more safe, we choose tracking error $\delta=0.125m$, and  obtain the following  condition on the minimum separation distance in the initial configuration:

\begin{equation}\label{eq:final_pmin}
    p_{min}\geq {\frac{2 \times (\delta+r)}{\epsilon_{min}}}={\frac{0.45}{\epsilon_{min}}}
\end{equation}
For our experiments, we chose $T = 30s$ as the total time for $2$D and $3$D continuum deformation coordination.

\subsubsection{Results of $2$D Continuum Deformation Coordination Experiment:}

\begin{table}[ht]
\begin{center}
\begin{tabular}{ |p{1.5cm}|p{1.5cm}|p{1.5cm}| } 
\hline
\hfil cf & \hfil 2D &  \hfil3D \\
\hline
\hfil $\epsilon_{1,f}$ & \hfil 1.8 & \hfil 0.9 \\ 
\hfil $\epsilon_{2,f}$ & \hfil 0.8 & \hfil 1.1 \\ 
\hfil $\epsilon_{3,f}$ & \hfil 1 & \hfil 0.7 \\ 
\hfil $\phi_{d,f}$ & \hfil 0 & \hfil 0.1 \\ 
\hfil $\theta_{d,f}$ & \hfil 0 & \hfil 0.12 \\ 
\hfil $\psi_{d,f}$ & \hfil -0.2 & \hfil 0.15\\ 
\hline
\end{tabular}
\end{center}
\caption{Final values of shear deformation angles and principal strains used for running simulations and experiments.}
\label{tab:configs}
\end{table}

\begin{table}[ht]
\begin{center}
  \begin{tabular}{|p{1.5cm}|p{1.5cm}|p{1.5cm}|p{1.5cm}|p{1.5cm}|p{1.5cm}|}
    \hline
    \multirow{2}{*}{} &
      \multicolumn{2}{c|}{2D (in $m$)} &
      \multicolumn{3}{c|}{3D (in $m$)} \\
      \hline
    \hfil CF & \hfil $x_i(0)$ & \hfil $y_i(0)$ & \hfil $x_i(0)$ & \hfil $y_i(0)$ & \hfil $z_i(0)$ \\
    \hline
    \hfil 1 & \hfil 1.00 & \hfil 0.50 & \hfil 2.25 & \hfil 2.25 & \hfil 1.50 \\
    \hfil 2 & \hfil 1.00 & \hfil 4.00 & \hfil 1.00 & \hfil 1.00 & \hfil 0.50 \\
    \hfil 3 & \hfil 3.00 & \hfil 2.25 & \hfil 1.00 & \hfil 4.00 & \hfil 0.50 \\
    \hfil 4 & \hfil 1.50 & \hfil 1.50 & \hfil 4.50 & \hfil 2.25 & \hfil 0.50 \\
    \hfil 5 & \hfil 1.50 & \hfil 3.00 & \hfil 1.50 & \hfil 1.50 & \hfil 0.75 \\
    \hfil 6 & \hfil 2.00 & \hfil 2.25 & \hfil 1.50 & \hfil 3.50 & \hfil 0.75 \\
    \hfil 7 & \hfil \textemdash & \hfil \textemdash & \hfil 3.50 & \hfil 2.25 & \hfil 0.75 \\
    \hfil 8 & \hfil \textemdash & \hfil \textemdash & \hfil 2.80 & \hfil 2.25 & \hfil 1.00 \\
    \hline
  \end{tabular}
\end{center}
  \caption{Initial positions of Crazyflies at $t=0s$}
  \label{table: init_loc}
\end{table}

\begin{figure}[ht]
\centering
\subfigure[]{\includegraphics[width=0.49\linewidth]{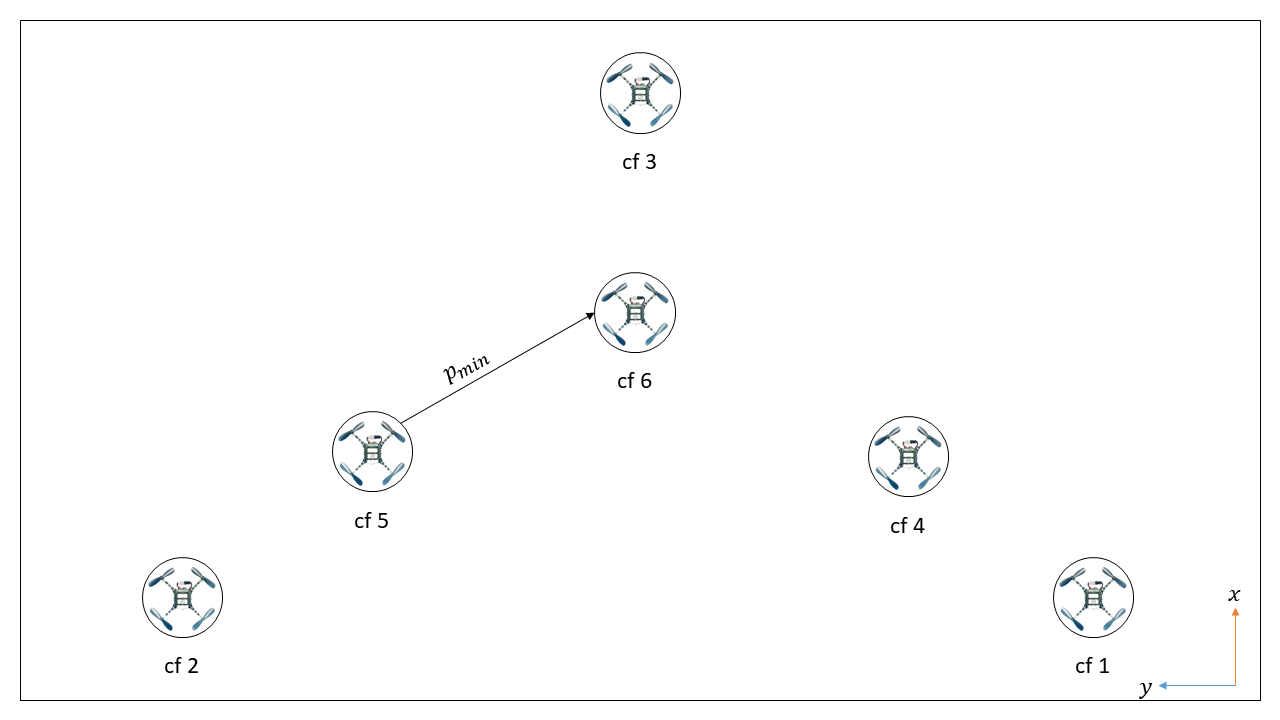}}
\subfigure[]{\includegraphics[width=0.49\linewidth]{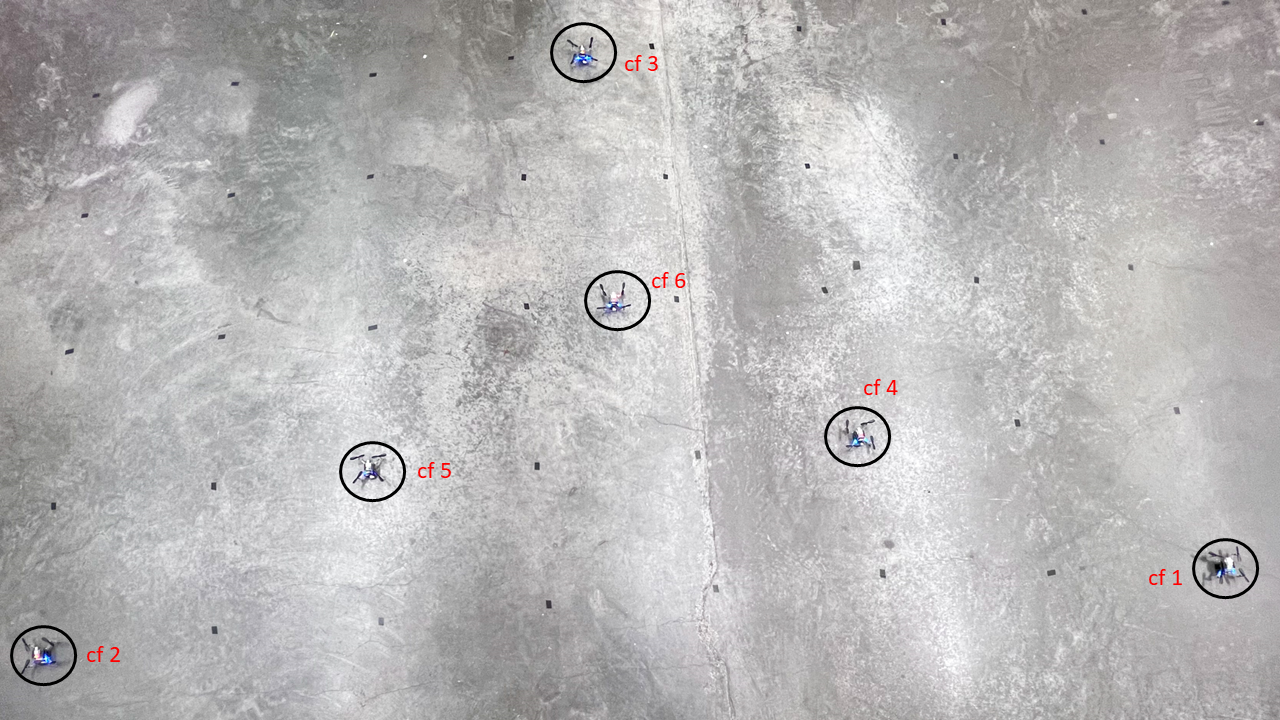}}
\caption{(a) Schematic and (b) actual formations of the quadcopter team in a $2$-D continuum deformation coordination. It can be seen that quadcopters $5-6$ and $4-6$, have the minimum separation distance $p_{min}=0.90 m$ in the initial configuration.}
\label{fig:2d_formation}
\end{figure}

\begin{figure}[ht]
\centering
\subfigure[]{\includegraphics[width=.49\linewidth]{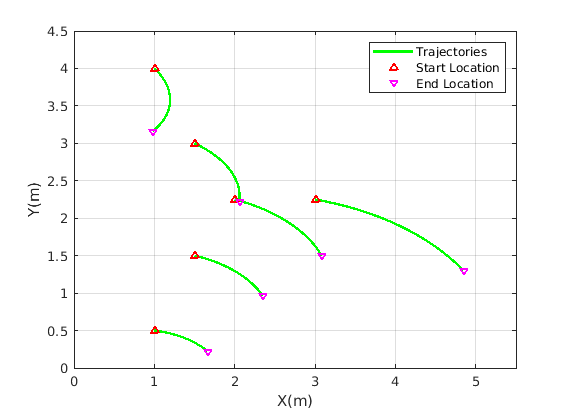}}
\subfigure[]{\includegraphics[width=.49\linewidth]{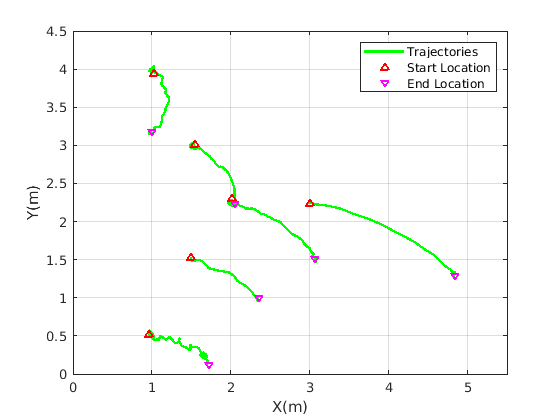}}
\caption{(a) Simulated paths of the quadcopter team in the $2$-D continuum deformation coordination. (a) The actual paths of the quadcopter team in the $2$-D continuum deformation coordination experiment.}
\label{fig:2d}
\end{figure}

According to Table \ref{tab:configs}, $\epsilon_{min} =0.8$ is considered for the $2$-D continuum deformation experiment. Substituting the value of $\epsilon_{min}$ in Eq. \eqref{eq:final_pmin}, we get $p_{min} = 0.5625 m$. As listed in Table \ref{table: init_loc}, cf $4-6$ and cf $5-6$ are equidistant and closest compared to any other pair of quadcopters. The distance between them is $0.90 m$ which satisfies Eq. \eqref{eq:final_pmin}, thus guaranteeing collision avoidance. The initial formation used for 2D experiments is shown in Figure \ref{fig:2d_formation}. Figure \ref{fig:2d}(a) shows the desired trajectories obtained via MATLAB whereas Figure \ref{fig:2d}(b) shows the observed trajectories in our experiments.

\begin{figure}[ht]
\centering
\subfigure[$\epsilon_1(0)=1$, $\epsilon_2(0)=1$, and $\epsilon_3(0)=1$]{\includegraphics[width=0.32\linewidth]{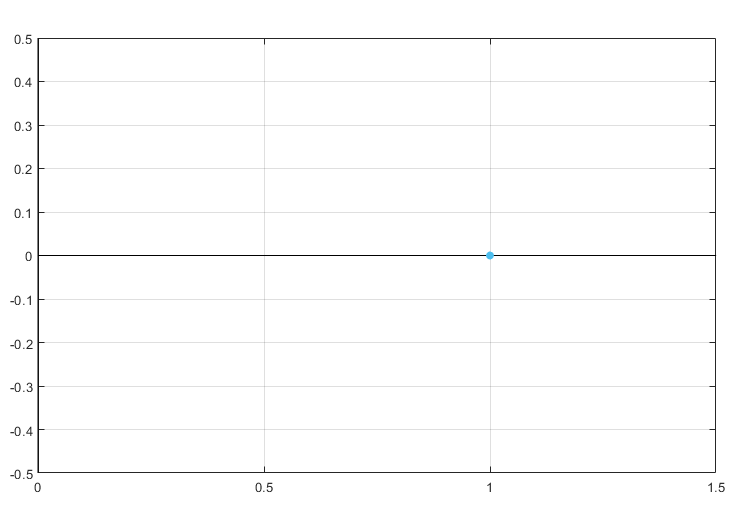}}
\subfigure[$\epsilon_1(10)=1.02$, $\epsilon_2(10)=0.98$, and $\epsilon_3(10)=0.94$]{\includegraphics[width=0.32\linewidth]{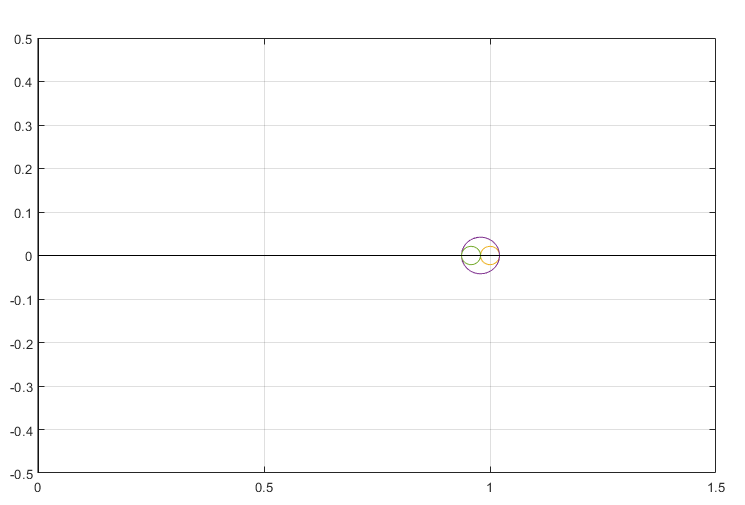}}
\subfigure[$\epsilon_1(15)=1.05$, $\epsilon_2(15)=0.95$, and $\epsilon_3(15)=0.85$]{\includegraphics[width=0.32\linewidth]{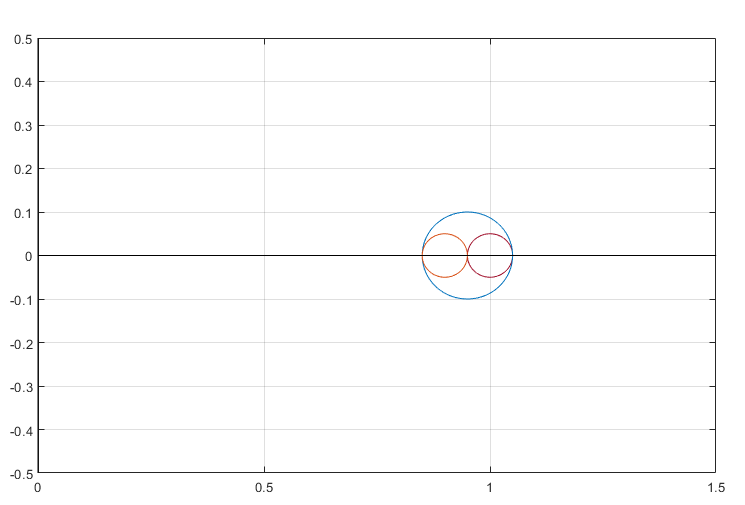}}
\subfigure[$\epsilon_1(20)=1.08$, $\epsilon_2(20)=0.92$, and $\epsilon_3(20)=0.76$]{\includegraphics[width=0.32\linewidth]{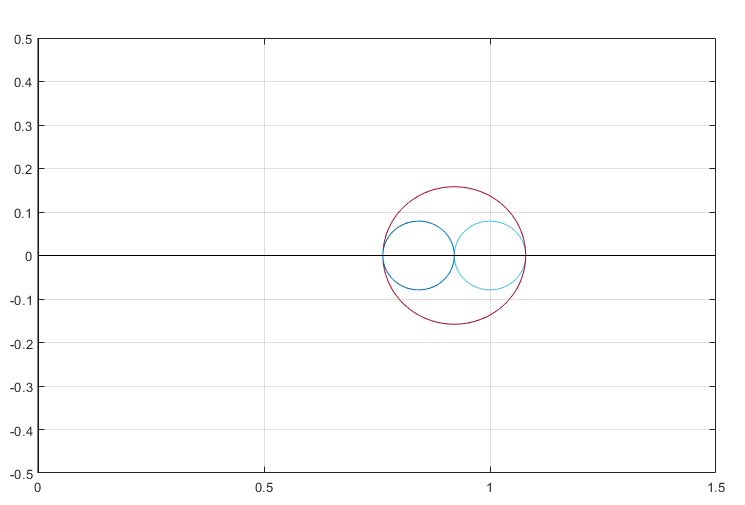}}
\subfigure[$\epsilon_1(25)=1.09$, $\epsilon_2(25)=0.9035$, and $\epsilon_3(25)=0.71$]{\includegraphics[width=0.32\linewidth]{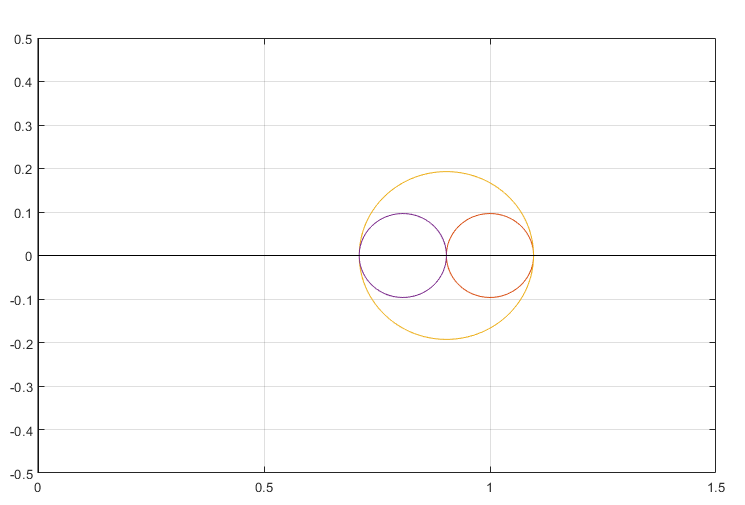}}
\subfigure[$\epsilon_1(30)=1.1$, $\epsilon_2(30)=0.90$, and $\epsilon_3(30)=0.70$]{\includegraphics[width=0.32\linewidth]{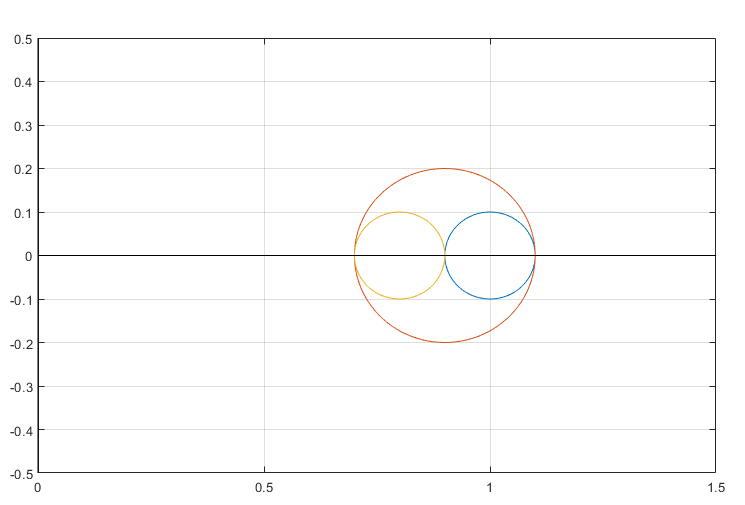}}
\caption{Mohr circle evolution: (a) $t=0s$, (b) $t=10s$, (c) $t=15s$, (d) $t=20s$, (e) $t=25s$, and (f) $t=30s$. 
}
\label{Mohrcircleevolution}
\end{figure}

\subsubsection{Results of $3$-D Continuum Deformation Coordination Experiment:}

\begin{figure}[ht]
\centering
\subfigure[]{\includegraphics[height=5cm, width=.49\linewidth]{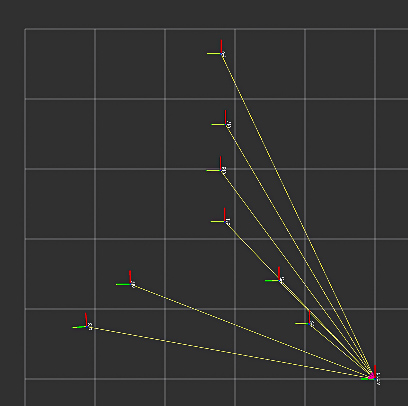}}
\subfigure[]{\includegraphics[height=5cm, width=.49\linewidth]{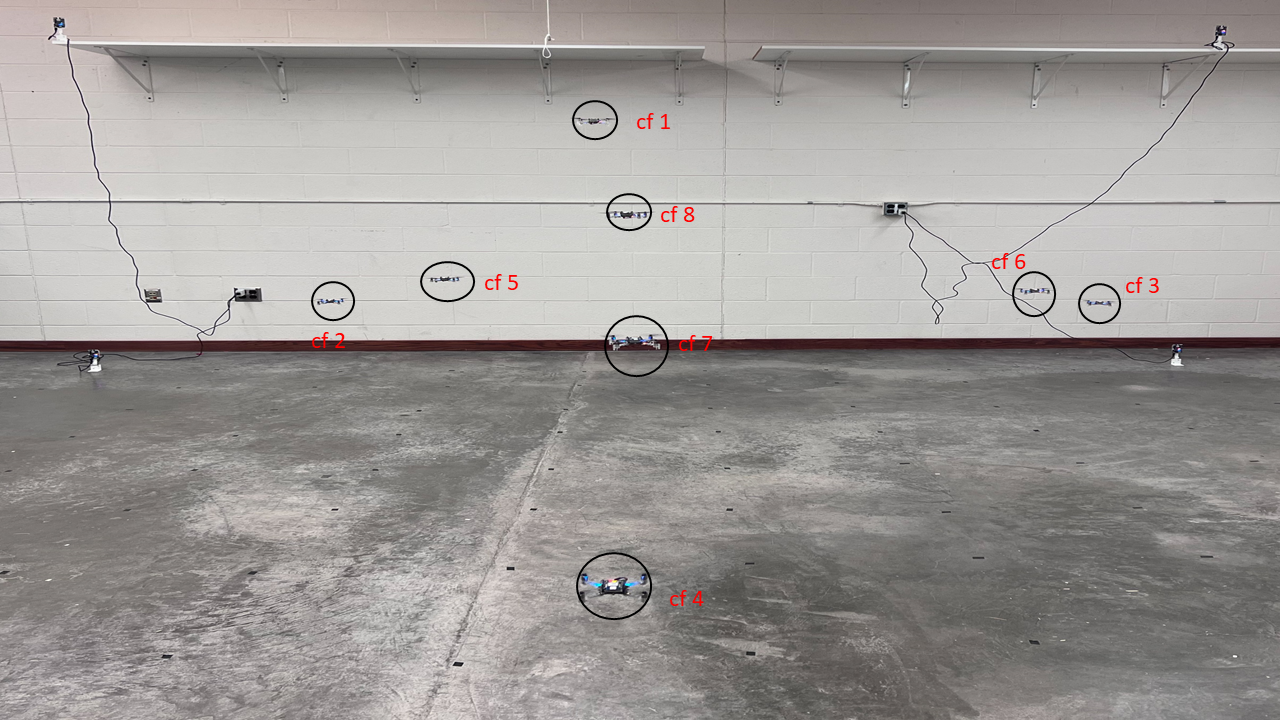}}
\caption{(a) Schematic and (b) actual formation of the quadcopter team in a $3$-D continuum deformation coordination. It can be seen that quadcopters $5-6$ and $4-6$, have the minimum separation distance $p_{min}=0.90 m$ in the initial configuration.}
\label{fig:3d_formation}
\end{figure}

\begin{figure}[ht]
\centering
\subfigure[]{\includegraphics[height=7cm, width=.49\linewidth]{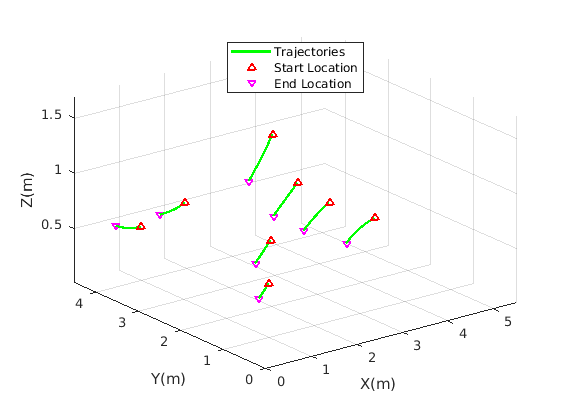}}
\subfigure[]{\includegraphics[height=7cm, width=.49\linewidth]{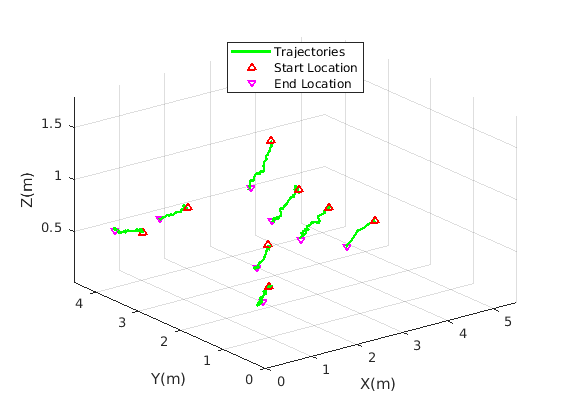}}
\caption{(a) Desired trajectories of the quadcopter team in $3$D continuum deformation coordination. (b) Actual trajectories of the quadcopter team in the $3$D continuum deformation coordination experiment.}
\label{fig:3d}
\end{figure}

According to Table \ref{tab:configs}, we have $\epsilon_{min} =0.7$. Substituting the value of $\epsilon_{min}$ in Eq. \eqref{eq:final_pmin}, we get $p_{min} = 0.64 m$. Similarly to 2D experiments, we observe from Table 1 that cf $1$ and cf $8$ are closest compared to any other pair of quadcopters. The distance between them is $0.74 m$ which satisfies Eq. \eqref{eq:final_pmin}. Figure \ref{Mohrcircleevolution} illustrates evolution of the Mohr circle,  characterizing the deformation of the quadcopters, at time $t=0s$, $t=10s$, $t=15s$, $t=20s$, $t=25s$, and $t=30s$.  Figure \ref{fig:3d_formation} shows the initial formation for $3$D continuum deformation coordination. Figure \ref{fig:3d}(a) shows the desired trajectories obtained via MATLAB. Figure \ref{fig:3d}(b) shows the final trajectories obtained in our experiments.

\section{Conclusions}\label{Conclusion}
In this paper, we treated quadcopters as particles of a deformable body and applied the principles of continuum mechanics to experimentally demonstrate the concept of linear deformation, principal strains, and shear strains. The primary objective of this work was to offer a new approach for teaching core concepts of mechanics by using quadcopters which can potentially provide valid multi-disciplinary research opportunities for the undergraduate students studying Engineering majors. Our secondary objective was to  experimentally validate continuum deformation guidance protocol and show how  a quadcopter team can aggressively deform but assure collision avoidance when they pass through narrow passages. 

\newpage

\bibliographystyle{spbasic}
\bibliography{reference}
\end{document}